\documentclass[prb]{revtex4}

\usepackage{graphicx}

\newcommand{\beq}{\begin{equation}}
\newcommand{\eeq}{\end{equation}}
\newcommand{\beqa}{\begin{eqnarray}}
\newcommand{\eeqa}{\end{eqnarray}}

\newcommand{\mnras}{Mon.\ Not.\ Royal Astron.\ Soc.\ }

\begin{document}

\title{The kinematic origin of the cosmological redshift}
\author{Emory F. Bunn} 
\email{ebunn@richmond.edu}
\affiliation{Department of Physics, University of Richmond, Richmond, Virginia 23173}
\author{David W. Hogg}
\affiliation{Center for Cosmology and Particle Physics and 
Department of Physics, New York University, 4 Washington Place, New
York, New
York 10003}

%\date{\today}

\begin{abstract}
A common belief about big-bang cosmology is that the cosmological redshift
cannot be properly viewed as a Doppler shift (that is, as evidence 
for a recession velocity), but must be
viewed in terms of the stretching of space. We argue that, contrary
to this view, the most
natural interpretation of the redshift is as a Doppler shift,
or rather as the accumulation of many infinitesimal Doppler shifts.
The stretching-of-space interpretation obscures a central idea of
relativity, namely that it is always valid to choose a coordinate system that is locally
Minkowskian.
We
show that an observed frequency shift in any spacetime can be
interpreted either as a kinematic (Doppler) shift or a gravitational
shift by imagining a suitable family of observers along the photon's path. 
In the context of the expanding universe the kinematic
interpretation corresponds to a family of comoving observers and
hence is more natural.
\end{abstract}

\maketitle

\section{Introduction}
\label{sec:intro}

Many descriptions of big-bang cosmology declare that the
observed
redshift
of distant galaxies is not a Doppler shift but is due to the
``stretching of space.'' The purpose of this paper is to examine the
meaning of such statements and to assess their validity. We wish to
make clear at the outset that we are not suggesting any doubt about
either the observations or the general-relativistic equations that
successfully explain them. Rather, our focus is on the
interpretation: given that a photon does not arrive at the observer
conveniently labeled ``Doppler shift,'' ``gravitational shift,'' or
``stretching of space,'' when can or should we apply these labels?

Arguably an enlightened cosmologist never asks this question.
In the curved spacetime of general relativity, there is no
unique
way to compare vectors at widely separated spacetime points,
and hence the notion of the relative velocity of a distant galaxy
is almost meaningless.
Indeed, the inability to compare vectors at different
points is the definition of a curved 
spacetime.\cite{baezbunn,carroll,schutz,mtw}
In practice, however, the enlightened view is far from universal.
The view presented by many cosmologists and astrophysicists,
particularly when talking to nonspecialists,
is that distant galaxies are ``really'' at rest,
and that the observed redshift is a consequence of some
sort of ``stretching of space,'' which is
distinct from the usual kinematic Doppler shift. In these descriptions,
statements that are artifacts of a particular coordinate system are
presented as if they were statements about the universe, resulting in
misunderstandings about the nature of spacetime in relativity.

In this paper we will show that the redshifts of distant objects in
the expanding universe
may be viewed as kinematic shifts due to relative
velocities, and we will argue that if we are forced to interpret
the redshift, this interpretation is
more natural than any other.

We begin with examples of the description of the cosmological redshift
in the first three
introductory astronomy textbooks chosen at random from the bookshelf of 
one of the authors.
\begin{itemize}
\item The cosmological redshift ``is {\it not} the same as a Doppler shift.
Doppler shifts are caused by an object's {\it motion through space},
whereas a cosmological redshift is caused by the {\it expansion of space}.''\cite{kaufmann} (Emphasis in original.)

\item ``A more accurate view [than
the Doppler effect] of the redshifts of galaxies is that the
waves are stretched by the stretching of space they travel through \ldots\ If space is stretching during all the time the light is traveling, the light
waves will be stretched as well.''\cite{fraknoi}

\item ``Astronomers often express redshifts as if they were radial velocities,
but the redshifts of the galaxies are not Doppler shifts \ldots\ Einstein's
relativistic Doppler formula applies to motion through space, so it
does not apply to the recession of the galaxies.''\cite{seeds}
\end{itemize}

More advanced textbooks often avoid this language. For instance,
the books by Peacock\cite{peacock} and Linder\cite{linder} give
particularly careful and clear descriptions of the nature of the
cosmological redshift. However, 
statements similar to those we have cited can be found even 
in some advanced textbooks. For example,
a leading advanced undergraduate level text
states that Doppler shifts 
``are produced by peculiar and not by recession velocities.''\cite{harrison}
In this paper we argue, as others have before 
us,\cite{chodorowskimilne,chodorowski,peacocknotes,whiting} 
that statements such as these are misleading
and foster misunderstandings about the nature of space and time. 

In general relativity the ``stretching of space''
explanation of the redshift is quite problematic. Light is 
governed by Maxwell's equations (or their general relativistic
generalization), which contain no ``stretching of space term'' and no
information on the current size of the universe. On the contrary, one
of the most important ideas of general relativity is that spacetime is
always locally indistinguishable from the (non-stretching) spacetime
of special relativity, which means that a photon doesn't know about the
changing scale factor of the universe.\cite{footnote1}

The emphasis in many textbooks on the stretching-of-spacetime
interpretation of the cosmological redshift 
causes readers to take too seriously the stretching-rubber-sheet
analogy for the expanding universe.
For example, it is sometimes stated as if it were obvious that ``it follows that all wavelengths of the light
ray are doubled'' if the scale factor doubles.\cite{harrison} Although
this statement is correct, it is not obvious. After all,
solutions to the Schr\"odinger equation, such as the electron orbitals
in the hydrogen atom, don't stretch as the universe
expands, so why do solutions to
Maxwell's equations? 

A student presented with the
stretching-of-space description of the redshift cannot be faulted for
concluding, incorrectly, that hydrogen atoms, the Solar System, and
the Milky Way Galaxy must all constantly ``resist the temptation'' to
expand along with the universe. One way to see that this belief is
in error is to consider the problem sometimes known as the ``tethered
galaxy problem,''\cite{harrisontethered,davistethered}
in which a galaxy is tethered to the Milky Way, forcing the distance
between the two to remain constant. When the tether is cut, does
the galaxy join up with the Hubble flow and start to 
recede due to the expansion of the universe?
The intuition that says that objects suffer from a temptation to 
be swept up in the expansion of the universe will lead to an affirmative
answer,
but the truth is the reverse:
unless there is a large cosmological constant
and the galaxy's distance is comparable to the Hubble 
length, the galaxy falls toward us.\cite{whiting,peacocknotes}
Similarly, it is commonly believed that the Solar System has a very
slight tendency to expand due to the Hubble expansion (although
this tendency is generally thought to 
be negligible in practice). Again, explicit calculation shows 
this belief 
not to be 
correct.\cite{sereno,cooperstock} The tendency to expand due 
to the stretching of space is nonexistent, not merely negligible.

The expanding 
rubber sheet is quite similar to the ether of pre-relativity physics in that 
although it is intuitively appealing, it makes no correct testable predictions,
and some incorrect ones such as the examples we have given.
It therefore has no rightful place in the theory.
(Some authors\cite{barnes,francis} 
have 
argued that considerations such as these do not refute the 
notion that space is really expanding. We agree with the calculations
in these papers but differ regarding the most
useful language to use to describe the relevant phenomena.)

In one set of circumstances the proper interpretation of the redshift
seems clear. When the curvature of spacetime is small over the
distance and time scales traveled by a photon, it is natural to
interpret the observed frequency shift as a Doppler shift. This interpretation is the reason that a police officer can give you a speeding ticket
based on the reading on a radar gun. As far as we know, no one has
successfully argued in traffic court that there is an ambiguity in
interpreting the observed frequency shift as a Doppler
shift.\cite{footnote3} In the expanding universe, spacetime curvature
is small over regions encompassing nearby objects, specifically
those with $z=\Delta\lambda/\lambda\ll 1$.
There should be no hesitation about calling the
observed redshifts Doppler shifts in this case, just as there is none
in traffic court. Surprisingly, however, many people seem to
believe that the ``stretching of space'' interpretation of the
redshift is the only valid one, even in this limit. We will examine
the interpretation of redshifts of nearby objects more carefully in
Sec.~\ref{sec:lowz}.

Aside from low-redshift sources, there is another case in which
spacetime curvature can be neglected in considering cosmological
redshifts, namely low-density cosmological models. An expanding universe
with density $\Omega=0$ (often known as the Milne model\cite{milne}) 
is merely the flat Minkowski spacetime of
special relativity expressed in nonstandard 
coordinates.\cite{footnote4}
In an $\Omega=0$ universe
there are no gravitational effects at all, so any
observed redshift, even of a very distant galaxy, must be a Doppler
shift. Furthermore, for low but nonzero density ($\Omega \ll 1$), 
the length scale associated with spacetime
curvature is much longer than the horizon distance.
Hence, spacetime curvature effects (that is, gravitational
effects) are weak throughout the observable volume, and the
special-relativistic 
Doppler shift interpretation remains valid even for galaxies with 
arbitrarily high
redshifts.

The more interesting cases are when $z$ and $\Omega$ are not small; that is, when the source is sufficiently distant that gravitational effects
are important over the photon's trajectory. The consensus 
is that the Doppler shift language must be eschewed
in this setting.
In Sec.~\ref{sec:highz} we review a standard 
argument\cite{peacock,peacocknotes}
that even in this case the redshift can be interpreted as the accumulation of
infinitesimal Doppler shifts along the line of sight, and we further argue
that there is a natural way to interpret the redshift as a single
(non-infinitesimal) Doppler shift.

A common objection to this claim is that the coordinate
velocity 
is not related to the redshift in accordance with the
special-relativistic Doppler formula.\cite{davissuperluminal,davisconfusion} 
However, the 
velocity referred to in this claim
is a mere artifact of a particular choice of time coordinate.
Specifically,
it is the rate of change of the distance to the object with respect to
the cosmic time coordinate, as measured at the present cosmic time.
This coordinate velocity is an unnatural quantity to discuss, because it depends on 
data outside of the observer's light cone.
The more natural velocity is the velocity of
the object at the time it crossed our past light cone, relative to
us today. This velocity is also a coordinate-dependent concept, 
but as we will show in Sec.~\ref{sec:highz}, the most natural way
to specify it operationally\cite{synge,narlikar} leads to a result
that is consistent with the special-relativistic Doppler formula.

In Sec.~\ref{sec:observers} we widen our focus to
consider frequency shifts in arbitrary curved spacetimes.
In any curved spacetime the observed frequency shift in a
photon can be interpreted as either a kinematic effect (a Doppler
shift) or as a gravitational shift. The two interpretations arise
from different choices of coordinates, or equivalently from imagining
different families of observers along the photon's path. We will
describe this construction explicitly, and show that
the comoving observers who are usually used to describe
phenomena in the expanding universe are the ones that
correspond to the Doppler shift interpretation.

\section{Redshifts of nearby galaxies are Doppler shifts}
\label{sec:lowz}

We begin by returning to the parable of the speeding ticket, 
mentioned in Sec.~\ref{sec:intro}.
\begin{quote}
``A driver is pulled over for speeding. The police officer says to the driver,
`According to the Doppler shift of the radar signal I bounced off your
car, you were traveling faster than the speed limit.'

``The driver replies, `In certain coordinate systems, the
distance between us remained constant during the time
the radar signal was propagating. In such a 
coordinate system, our relative velocity is zero, and
the observed wavelength shift was not a Doppler shift. So
you can't give me a ticket.' ''
\end{quote}

If you believe that the driver has a legitimate argument, then you
have our permission to believe that cosmological redshifts are not
really Doppler shifts. If, on the other hand, you think that the officer
is right, and the redshift can legitimately be interpreted as a Doppler shift,
then you should believe the same thing about redshifts of nearby galaxies
in the expanding universe.

Why is the police officer right and the driver wrong? 
Assuming the officer majored in physics, he might
explain the situation like this: ``Spacetime in my neighborhood is very
close to flat. 
That means that I can lay down space and time
coordinates in my neighborhood such that, to an excellent
approximation, the rules of special relativity hold. 
Using those
coordinates, I can interpret the observed redshift as a Doppler shift
(because there is no gravitational redshift in flat spacetime) and
calculate your coordinate velocity relative to me. The errors
in this method are of the same order as the departures from flatness
in the spacetime in a neighborhood containing both me and you. 
As long as I'm willing to put up
with that very small level of inaccuracy, I can interpret that coordinate
velocity as your actual velocity relative to me.''

The principle underlying the officer's reasoning is uncontroversial. It is no different
from the principle that lets football referees ignore
the curvature of Earth and use a flat coordinate
grid in describing a football field.

We now return to the consideration
of redshifts in the expanding universe.
As noted in Sec.~\ref{sec:intro},
it is instructive in this context (as in many others) to start with
the zero-density expanding universe (the Milne 
model).\cite{chodorowskimilne}
The Milne universe consists of
a set of galaxies expanding outward from 
an initial Big Bang, with the galaxies assumed to have negligible
mass, so that the geometry is that of gravity-free Minkowski
spacetime. This spacetime can be described either
in the usual comoving coordinates
of cosmology or in Minkowskian coordinates. 
Because there is no
spacetime curvature and no gravity in this universe, it is clear that
the observed redshifts should be interpreted as Doppler shifts.
In comoving coordinates, however, the redshifts are easily seen to be 
the usual cosmological redshift. In this case there is
no distinction between the cosmological redshift and the Doppler shift.

In our actual universe spacetime is not exactly flat, but
we can approximate it as flat in a small neighborhood.
It might be tempting to think that ``approximating away'' the curvature
of spacetime is the same as approximating away the expansion altogether.
(It seems to us that people who believe that cosmological redshifts
cannot be viewed as Doppler shifts, even arbitrarily nearby,
often believe something like this.)
However, this belief is incorrect. 
When we approximate a small neighborhood
of an expanding spacetime as flat, we make errors of order $(r/R_c)^2$
in the metric, where $r$ is the size of the neighborhood and $R_c$ is the 
curvature scale (generally the Hubble length). The redshifts of 
galaxies in that neighborhood are of order $(r/R_c)$, so they are not
approximated away in this limit. 
(The Milne model corresponds to the limit $R_c\to\infty$,
in which case no approximation is made.)

In comoving coordinates the
spacetime line element for the Robertson-Walker expanding universe
is
\beq
ds^2=-c^2dt^2+[a(t)]^2\Big(dr^2+[S(r)]^2[d\theta^2+\sin^2\theta\,d\phi^2]\Big).
\eeq
Here $S(r)=r$ for a flat universe. For a closed universe
with positive curvature $K$, $S(r)=K^{-1}\sin(Kr)$, and for
an open universe with negative curvature $-K$, $S(r)=K^{-1}\sinh(Kr)$.
For realistic models $K^{-1}$ is at least comparable to the Hubble
length, and thus $S(r)\approx r$ for nearby points in all cases.

Assume that an observer at the origin
at the present time $t_0$ measures the redshift of a 
galaxy at some comoving distance $r$. Assume that the
galaxy is nearby so that $r/R_c\ll 1$, where $R_c$ is the curvature
length scale.
Over such a distance scale spacetime can be well
approximated as flat. Let the observer lay down coordinates
that approximate spacetime in her neighborhood as flat as well as possible.
That is, let the observer choose coordinates $T,R,\theta,\phi$ such that
\beq
ds^2\approx -c^2dT^2+dR^2+R^2(d\theta^2+\sin^2\theta\,d\phi^2)
\eeq
with the smallest possible errors in her neighborhood. There
are several procedures for making this choice, such as choosing Riemann normal
coordinates. The resulting errors are of order
$(r/R_c)^2$ in the metric.
If we regard such errors as negligible, then 
we can legitimately approximate spacetime as flat. In such a coordinate
system a comoving galaxy has a velocity $v$ relative
to the observer, which is related to the observed redshift in the expected
way to 
leading order:
\beq
z = (v/c)+O((r/R_c)^2).
\eeq

\begin{figure}[t]
\includegraphics[width=2.5in]{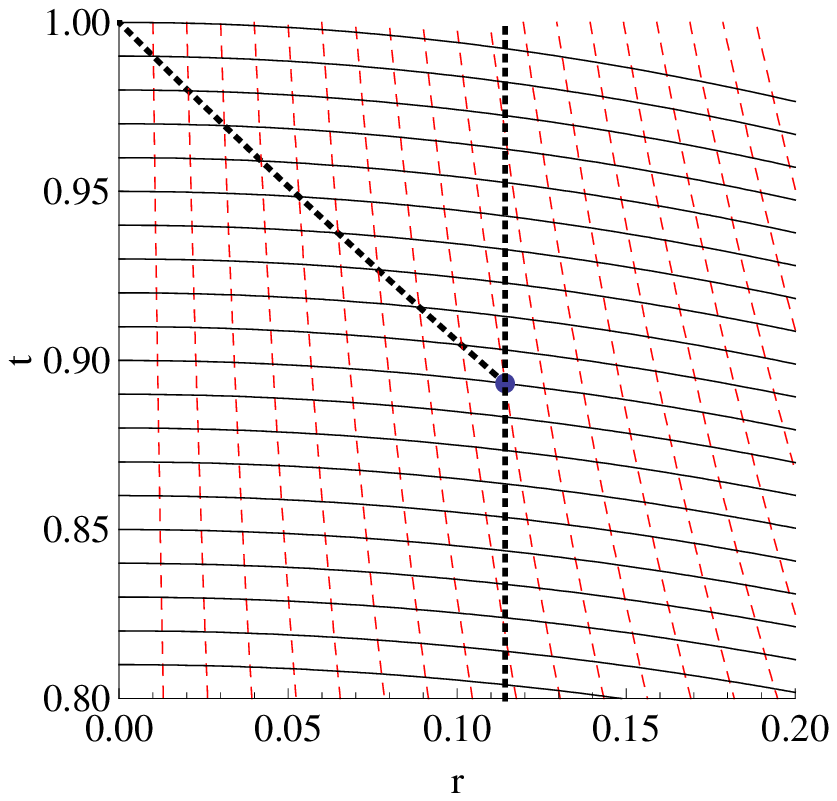}
\qquad\qquad\includegraphics[width=2.5in]{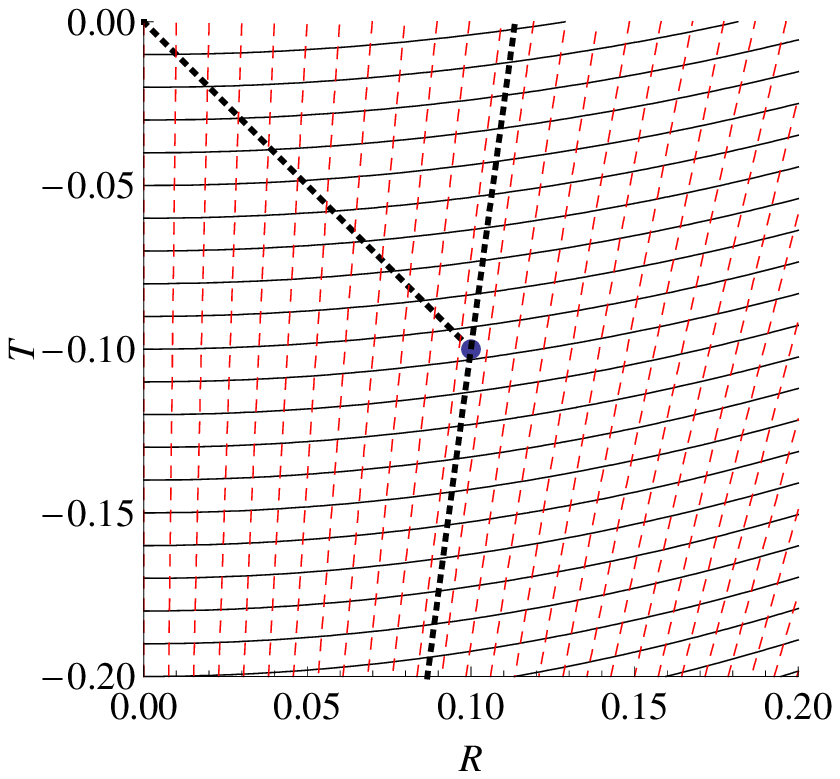}
\caption{Spacetime in different coordinate systems. (a) Plot of an expanding Robertson-Walker universe in the usual comoving
coordinates $(r,t)$. Coordinates are expressed in units of the
Hubble length and time. The observer is located at $r=0$. The dotted
curves show the world lines of a particular galaxy and of a light signal
from that galaxy reaching the observer at the present time ($t=1$). The
dashed and solid curves are contours of constant Riemann normal coordinates
$(R,T)$; that is, the coordinates that approximate flat spacetime
as well as possible near the observer. (b) The
roles of the two coordinate systems are reversed: the coordinate axes
are the Riemann normal coordinates $(R,T)$, and the dashed and solid curves
are contours of constant comoving coordinates $(r,t)$. As
time passes, the galaxy moves to larger values of $R$; that is, it is
not at rest in this coordinate system.}
\label{fig:riemannnormal}
\end{figure}

Figure \ref{fig:riemannnormal} illustrates the coordinate systems. In Fig.~\ref{fig:riemannnormal}(a) the horizontal and vertical axes are standard
comoving coordinates $(r,t)$. The solid and dashed curves
show contours of the coordinates $(R,T)$ that best approximate
spacetime as flat. World lines of a galaxy and
of a light signal from the galaxy to the observer are also plotted.
Figure~\ref{fig:riemannnormal}(b) shows the view of spacetime in which the
metric appears closest to flat spacetime; that is, it is
the one to use when using special relativistic language to
describe our cosmic neighborhood. In this coordinate system comoving
galaxies are moving away from the observer, and the observed redshift
is a Doppler shift.

If we assume that the observed galaxy is at a redshift much less than one, the 
errors in approximating spacetime to be flat are small. As long as those errors
are small enough to be neglected, the
observer is in the same situation as in the speeding ticket parable. We don't 
merely say that the police officer 
is \textit{allowed}
to regard the observed redshift as a Doppler shift. We say that that is 
\textit{the}
natural interpretation of the shift. The same statement
is true in the cosmological case. The only natural interpretation
of the redshift of a nearby galaxy is as 
a Doppler shift.

\section{\label{sec:highz}Redshifts of distant galaxies can be regarded as Doppler 
shifts}

For more distant galaxies, that is, those with redshifts of order one or more,
the light-travel distances involved are comparable to the curvature
length scale, and thus the Riemann normal coordinate system
cannot be used to approximate spacetime as flat with good accuracy. In this
regime we might imagine being required to drop the Doppler
interpretation in favor of the ``stretching of space'' point of view.
We now argue that the Doppler interpretation is valid even in this
regime.

Let us begin by reviewing
a standard derivation\cite{peacock,peacocknotes} 
of the cosmological redshift.
Consider a photon that travels from a galaxy to a distant observer,
both of whom are at rest in comoving coordinates. Imagine a family
of comoving observers along the photon's path, each of whom
measures the photon's frequency as it goes by. We assume
that each observer is close enough to his neighbor so that we can accommodate
them both in one inertial reference frame and use special
relativity to calculate the change in frequency from one observer
to the next. If adjacent observers are separated
by the small distance $\delta r$, then their relative speed in this
frame is 
$\delta v=H \delta r$,
where $H$ is the Hubble parameter.
This speed is much less than $c$, so the frequency shift is
given by the nonrelativistic Doppler formula
\beq
\delta\nu/\nu=-\delta v/c=-H \delta r/c=-H \delta t.
\eeq
We know that $H=\dot a/a$ where $a$ is the scale factor.
We conclude that the frequency change is given by
$\delta\nu/\nu=-\delta a/a$; that is, the frequency decreases in inverse proportion 
to the scale factor. The overall redshift is therefore given by
\beq
1+z\equiv{\nu(t_e)\over \nu(t_o)}={a(t_o)\over a(t_e)},
\label{eq:redshift}
\eeq
where $t_e$ and $t_o$ refer to the times of emission and observation, respectively.

In this derivation we interpret the redshift as the accumulated effect
of many small Doppler shifts along the photon's path. 
We now address the question of whether it makes sense to
interpret the redshift as one big Doppler shift, rather than
the sum of many small ones.

\begin{figure}[t]
\includegraphics{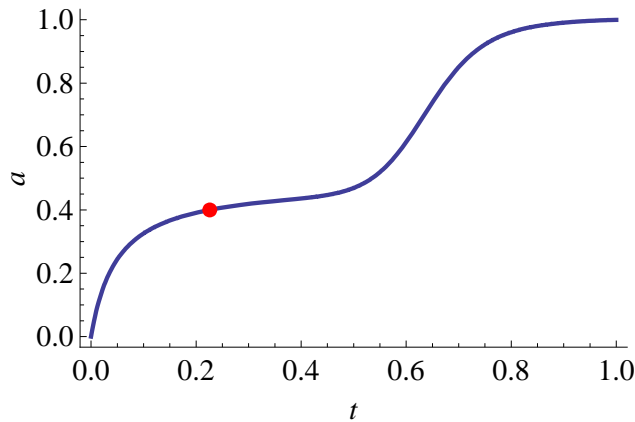}
\caption{The scale factor of a
hypothetical ``loitering'' universe as a function of time (measured
in units of the present time). At the time indicated by the dot, a
galaxy emits radiation, which is observed at the present time. At the
times of both emission and observation, the expansion is very slow,
yet the galaxy's observed redshift is large.}
\label{fig:loiter}
\end{figure}

Figure \ref{fig:loiter} shows a common argument against such an 
interpretation.
Imagine
a universe whose expansion rate varies with time as shown in Fig.~\ref{fig:loiter}.
A galaxy emits radiation at the time $t_e$ indicated by the dot, and the
radiation reaches an observer at the present time $t_0$. The observed redshift
is $z=a(t_0)/a(t_e)-1=1.5$, which by the special-relativistic
Doppler formula would correspond to a speed of $0.72c$. At the times of
emission and absorption of the radiation, the expansion rate is very slow,
and the speed $\dot a r$ of the galaxy is therefore much less than this value.
We can construct models in which both $\dot a(t_e)$ and $\dot a(t_0)$ are 
arbitrarily small without changing the ratio $a(t_0)/a(t_e)$ and hence
without changing the redshift.

Upon closer examination this argument is unconvincing,
because the calculated velocities are not the correct
velocities. We should not calculate velocities at a fixed instant of
cosmic time (either $t=t_e$ or $t=t_0$). Instead we should calculate the
velocity of the galaxy {\it at the time of light emission} relative to
the observer {\it at the present time}. After all, if
a distant galaxy's redshift is measured today, we wouldn't
expect the result to depend on
what the galaxy is doing today,
nor
on what the observer was doing long
before the age of the dinosaurs.

In fact, when astrophysicists talk about what a distant object
is doing ``now,'' they often do not mean at the present value of 
the cosmic time, but rather at the time the object crossed
our past light cone. For instance, 
when astronomers measure the orbital speeds of planets
orbiting other stars, the measured velocities are always
of this sort.
There is an excellent reason for this convention: we never have information
about what a distant object is doing (or if the object even exists) 
at the present cosmic time.
Any statement in which ``now'' 
is used to refer to the present cosmic time at the location of a distant
object is not about anything observable, because it refers
to events far outside our light cone.

In summary, 
if we wish to discuss the redshift of a distant galaxy as a Doppler
shift, we need to be willing to talk about $v_{\rm rel}$, 
the velocity of the galaxy
\textit{then} relative to us \textit{now}. 
Talking about $v_{\rm rel}$ is precisely the sort of thing that the
enlightened cosmologist described in Sec.~\ref{sec:intro} 
refuses to do because,
in a curved spacetime, there is not a unique way
to define the relative velocity of objects at widely separated spacetime events.
Determining the velocity of one object relative to another involves
comparing the two objects' velocity four-vectors. [To be specific,
the dot product of these vectors is 
$\gamma_{\rm rel}\equiv (1-v_{\rm rel}^2
/c^2)^{-1/2}$.] To do that
we have to transport one of the vectors to the location of the other.
In a curved spacetime, the result of such a ``parallel transport'' depends
on the path along which the vector is transported. 

Suppose that we wish to defy the purist and talk about $v_{\rm rel}$.
We have to parallel transport the galaxy's velocity
four-vector to the observer's location. As others
have noted before,\cite{synge,narlikar} the only natural path to choose
for this parallel transport is the path followed by the light
(that is, the null geodesic joining the emission and observation events). 
If we follow this procedure to determine the relative velocity,
we find that the velocity obeys the rule
\beq
\sqrt{c+v_{\rm rel}\over c-v_{\rm rel}}={a(t_o)\over a(t_e)}.
\label{eq:srdoppler}
\eeq
As noted in Eq.~(\ref{eq:redshift}), 
the ratio of scale factors is equal to 
$1+z$, so Eq.~(\ref{eq:srdoppler}) is the special-relativistic Doppler
formula. 
In other words, the relative speed
$v_{\rm rel}$ as defined by parallel transport is related
to the observed redshift as it should be if the redshift
is a Doppler shift.

\begin{figure}[t]
\includegraphics{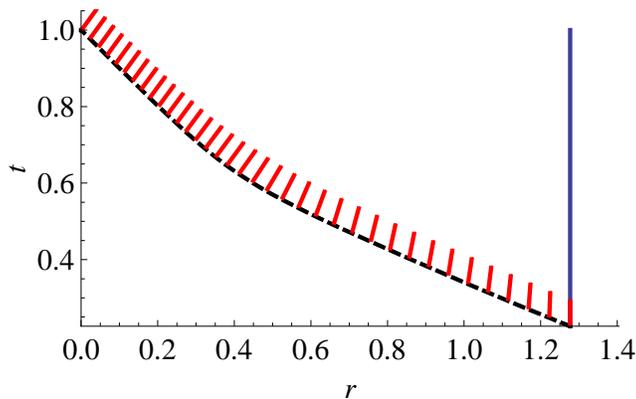}
\caption{Parallel transport of the velocity four-vector.
The solid and dashed
lines show the world line of a galaxy and the path of radiation
from the galaxy to the observer in comoving coordinates. 
The short solid lines show the 
galaxy's four-velocity being parallel transported to the observer along
the path of the light.}
\label{fig:transport}
\end{figure}

Figure \ref{fig:transport} illustrates this process. The solid line
is the world line of a galaxy, and the dashed line
is the path of light traveling from the
galaxy to the observer. The scale factor is as shown in Fig.~\ref{fig:loiter}. The short solid lines indicate the galaxy's velocity
four-vector as it is parallel transported along the light path to the observer.
During the times when the universe is expanding slowly
the direction of this vector doesn't change significantly, but it does
when the universe expands more rapidly. 

Equation (\ref{eq:srdoppler}) can be derived by a straightforward
calculation using the rules for parallel transport, but
the derivation is easier if we recast the
statement in more physical terms. 
As we did at the beginning of this section, 
imagine many comoving observers stationed along the line from the observed
galaxy to the observer. Each observer has a local reference frame in
which special relativity can be taken to apply, and the observers are 
close enough together that each one lies within the local frame of
his neighbor. Observer number 1, who is located near the original galaxy,
measures its speed $v_1$
relative to him and gives this information to observer 2.
Observer 2 measures the speed $u$ of observer 1 relative to her,
adds this speed
to the speed of the galaxy relative to observer 1 
using the usual special-relativistic formula, 
\begin{equation}
v_2={v_1+u\over 1+v_1u/c^2},
\end{equation}
and interprets the result
as $v_{\rm rel}$, 
the speed of the galaxy relative to her. She passes this information
on to the next observer, who follows the same procedure, as does
each subsequent observer. At each stage, the imputed velocity of the
original galaxy relative to the observer will match the
redshift of the galaxy in accordance with Eq.~(\ref{eq:srdoppler}).
The procedure we have described operationally is equivalent
to performing parallel transport of the galaxy's four-velocity and using
that vector at each stage to calculate $v_{\rm rel}$.

We can express this derivation in yet another way. Consider the ``world tube''
obtained by drawing a sphere with a small radius $\epsilon$ around
each point on the path of the light ray from the galaxy to the observer.
This region of spacetime can be considered as flat Minkowski spacetime
up to errors of order $\epsilon^2$. (This statement is true for any
small neighborhood around a geodesic.) Within this tube
the observed redshift can be explained only as a Doppler shift, because
the spacetime is flat to arbitrary precision. 
This argument
supplies the simplest proof we can think of that $v_{\rm rel}$ is
related to the redshift by Eq.~(\ref{eq:srdoppler}).

We do not expect to have convinced purists who refuse even to talk
about $v_{\rm rel}$ to change their minds. They would dismiss $v_{\rm
rel}$ as a mere ``coordinate velocity,'' not an ``actual velocity,''
and take no interest in it. If purists have the courage
of their convictions, they would say 
that it makes no sense to try to
attach labels such as ``Doppler'' or ``gravitational'' to the observed
redshift. This position is unassailable, and we have no wish to argue
against it. We do claim, however, that if you wish to try to
talk about $v_{\rm rel}$, then the definition proposed in this section
is the most natural way to do so.
Because this
definition of $v_{\rm rel}$ results in the Doppler formula entirely
explaining the redshift, we 
claim that you should either be a purist and refuse to try to label
the cosmological redshift, or you should label it a Doppler shift.

It is interesting to contrast the position we advocate with another
approach\cite{bondi,peacocknotes} that describes the cosmological
redshift as a particular combination of Doppler and gravitational
terms. In that approach comoving coordinates are used to specify the
velocity of the distant galaxy relative to the observer, and the
observed redshift is decomposed into a Doppler term based on this
velocity and a gravitational blueshift, which can be calculated by
considering the gravitational potential due to the matter in a sphere
centered on the observer. The difference between the two approaches
is in the way the velocity is defined. Because
there is no unique specification of relative velocity for distant
objects in curved spacetime, both approaches
are correct. We believe that the approach based
on defining $v_{\rm rel}$ through parallel transport is more natural
than the one based on a particular choice of coordinates.

\section{Kinematic and gravitational redshifts in general spacetimes}
\label{sec:observers}

Let us step back from the specific case of the expanding universe and
consider the propagation of a photon in an arbitrary spacetime. A
photon is emitted at some spacetime point ${\cal P}_e$ with a
frequency $\nu_e$ and is absorbed at ${\cal P}_a$ with frequency $\nu_a$.
The two frequencies are measured by observers in local inertial
reference frames at the two spacetime points. If there is
a frequency shift ($\nu_e\ne \nu_a$), under what circumstances should we
describe it as a Doppler shift or as a gravitational shift?

A natural way to answer this question is to populate the space
between the photon's emission and observation with a dense family of observers.
Let the first observer be
at rest relative to the emitter of the photon at the
emission event, and let the last be at rest relative
to the absorber at the absorption event. Finally,
let the velocities of 
observers vary smoothly along the photon's path. 
As in Sec.~\ref{sec:highz}, we can explain the
observed frequency shift by adding up the shifts as the photon
passes from one observer to the next. We can construct two 
such families
of observers: one in which each of the shifts is a Doppler shift,
and one in which each is a gravitational shift. 
By reference to these families, we can interpret the observed shift
as the accumulation of either many small Doppler shifts or many
small gravitational shifts.

The Doppler and gravitational families are not the
only possibilities. We can construct
other families with respect to whom
both Doppler and gravitational shifts would be seen to contribute.
As noted in Sec.~\ref{sec:highz}, for example,
one way of regarding the cosmological 
redshift\cite{bondi,peacocknotes} splits it into a specific combination
of Doppler and gravitational terms.
The Doppler and gravitational families provide
two of many different ways of interpreting the observed shift. In any given
situation, we can argue about which (if either) of the two
families is natural to consider.

To construct the Doppler family of observers, we require all
observers to be in free fall. In this case, within each local inertial
frame, there are no gravitational effects, and hence the infinitesimal
frequency shift from each observer to the next is a Doppler shift.

To construct the gravitational family of observers, we require that
each member be
at rest relative to her neighbor at the
moment the photon passes by, so that there are no Doppler
shifts. Initially, it might seem impossible in general
to satisfy this condition
simultaneously with the condition that the first and last observers 
be at rest relative to the emitter and absorber, but it is always
possible to do so. One way to see that it is possible is to draw a small world tube around
the photon path as in Sec.~\ref{sec:highz}. Within this tube,
spacetime is arbitrarily close to flat. We can construct ``Rindler
elevator coordinates,'' the special-relativistic generalization
of a frame moving with uniform acceleration, within this tube, such
that the velocities at the two ends match up correctly.\cite{rindler} 
The members of the gravitational family are at rest
in these coordinates.
Because they are not in free fall, the members of the gravitational
family all feel like they are in local gravitational fields. Because
each has zero velocity relative to her neighbor when the photon
goes by, each observer interprets the shift in the photon's frequency
relative to her neighbor as a gravitational shift. 

Because the two families exist for any photon path, we can always
describe any frequency shift as either Doppler or gravitational. 
In some situations, one seems clearly more natural than the other.
When we discuss the Pound-Rebka experiment,\cite{poundrebka} which measured
the redshift of photons moving upward in Earth's gravitational field,
we generally choose to regard observers fixed relative to the Earth
(the gravitational family) as more natural than free-fall
observers, and hence we interpret the measurement as a gravitational
redshift. In contrast, if you were falling past Pound and Rebka
in a freely falling elevator as they performed the experiment, you might
choose a Minkowskian inertial frame encompassing you and the
entire experiment. Within this frame, you
would,
by the equivalence principle, interpret their results as a Doppler shift.
In so doing, you would be choosing to regard the 
Doppler family as the natural one, because this family is the one whose
behavior is simplest to describe in your chosen frame.
Finally, if you tried to beat a speeding ticket
by claiming that the radar results were due to a gravitational
redshift, you would in effect be considering a ``gravitational family''
of prodigiously accelerating observers, with one at rest relative
to the radar gun and one at rest relative to the driver. Needless
to say, 
the police officer who gives you a ticket
regards this family as 
extremely unnatural.

In the cosmological context, we almost always work with
freely
falling comoving observers, a choice which corresponds precisely
to the Doppler family. This choice is natural because it respects
the symmetries of the spacetime and makes
the mathematical description as simple as possible. 
On the other hand, the gravitational family is so unnatural in cosmology that,
as far as we know, it has never been used for anything.
Because the Doppler family is by far the most natural family to
work with, it is natural to interpret the cosmological redshift
as a Doppler shift, and it is curious that this
interpretation is frequently frowned upon.

\section{Why the interpretation matters}

There is no ``fact of the matter'' about the interpretation of the 
cosmological redshift: what one concludes depends on one's coordinate
system or method of calculation.  Nonetheless,
it is instructive to analyze the
differing interpretations of the redshift, partly to
improve the understanding of cosmology, but more importantly to improve the
understanding of general relativity. That analysis leads us to conclude that the most natural interpretation of the redshift is kinematic.

One of the key ideas of
general relativity is the importance of distinguishing between
coordinate-independent and coordinate-dependent statements. Another
is the idea that spacetime is always locally indistinguishable from
Minkowski spacetime. 
Cosmology instructors, books (especially at the introductory level), 
and students often
fall into the fallacy of reifying the rubber sheet; that is,
treating the expanding-rubber-sheet model of space as if it were 
a real substance. This error leads people away from both of
these key ideas and causes mistaken intuitions such as that
the Milky Way
Galaxy must constantly ``resist the temptation'' to expand with
the expanding universe or that the ``tethered galaxy'' described
in Sec.~\ref{sec:intro} moves
away after the tether is cut.

The common belief that the cosmological redshift can only be explained
in terms of the stretching of space is based on conflating the
properties of a specific coordinate system with properties of space
itself. This confusion is precisely the opposite of the correct frame of mind
in which to understand relativity.

\begin{acknowledgments}
We thank Sean Carroll, John Peacock, Jim Peebles, Dave Spiegel, and Ned Wright
for helpful comments. This work was supported in part by NSF grant
AST-0507395 (EFB) and a research fellowship from the Alexander von
Humboldt Foundation of Germany (DWH).
\end{acknowledgments}


\begin{thebibliography}{10}
\bibitem{footnote1}
One might argue against this claim by noting that the general relativistic Maxwell equations contain covariant derivatives, which ``know about'' the expansion. However, covariant derivatives are locally indistinguishable from ordinary derivatives because a local Minkowski frame can always be chosen.
\bibitem{footnote3}
If you try this argument, let us know how it works out for you.
\bibitem{footnote4}
This fact is an occasional source of confusion, because a low-density universe is an open universe, in which space, as opposed to spacetime, has negative curvature. What matters in the present context is the spacetime curvature, that is, the deviation of the overall spacetime metric from that of special relativity, not the spatial curvature. The latter describes the geometry of a ``slice'' through the spacetime at constant time in a particular coordinate system.
\end{thebibliography}
\end{document}